\documentclass{PoS}
\usepackage{colordvi,psfig,epsfig}

\PoS{PoS(LAT2005)228}

\title{Two-loop evaluation of large Wilson loops with overlap fermions: 
the b-quark mass shift, and the quark-antiquark potential}

\ShortTitle{Two-loop evaluation of large Wilson loops with overlap fermions}

\author{\speaker{Andreas Athenodorou}\\
        Department of Physics, University of
        Cyprus, Nicosia CY-1678, Cyprus\\
        E-mail: \email{ph00aa1@ucy.ac.cy}}

\author{Haralambos Panagopoulos\\
        Department of Physics, University of
        Cyprus, Nicosia CY-1678, Cyprus\\
        E-mail: \email{haris@ucy.ac.cy}}

\abstract{We compute, to two loops in pertubation theory, the fermionic contribution to
 rectangular $R\times T$ Wilson loops, for different values of $R$ and $T$.

We use the overlap fermionic action. We also employ the clover action,
 for comparison with existing results in the literature.
 In the limit $R, T \rightarrow \infty$ our results lead to the shift in the b-quark mass.
We also evaluate the perturbative static potential as $T \rightarrow \infty$.}

\FullConference{XXIIIrd International Symposium on Lattice Field
Theory\\
		 25-30 July 2005\\
		 Trinity College, Dublin, Ireland}

\begin{document}

\section{Introduction}

In this work, we compute the perturbative value of \WildStrawberry{{\bf large Wilson loops}} 
  up to two loops, using the \RedViolet{clover} (SW) and \RedViolet{overlap} fermions. 
Using the perturbative values of Wilson loops 
of infinite length, we evaluate the shift of the \WildStrawberry{{\bf b-quark mass}}. 
The perturbative values of Wilson loops 
of infinite time extent lead us to the evaluation of the \WildStrawberry{{\bf static potential}}. 
For the case of clover fermions, we also compare our results with established results. 

The calculation of Wilson loops 
in lattice perturbation theory is useful in a number of ways:
a)~It leads to the prediction of a strong coupling constant 
\Blue{$a_{\overline{MS}}\left( m_Z \right)$} from 
low energy hadronic phenomenology by means of non-perturbative 
lattice simulations (for a list of relevant references, see our longer write-up~\cite{AAPH}).
b)~It is employed in the context of mean field improvement programmes 
of the lattice action and operators.
c)~In the limit of infinite time separation, 
$T \rightarrow \infty$, Wilson loops give access to the perturbative 
quark-antiquark potential. Furthermore in the limit of large distances, 
$R \rightarrow \infty$, the self energy of static sources can be 
obtained from the potential, enabling the calculation of 
 \Blue{$\overline{m_b}\left(\overline{m_b}\right)$} from non-perturbative 
simulations of heavy-light mesons in the static limit~\cite{GMCTS}.

\section{Calculation of Wilson loops}

The Wilson loop $W$, around a closed curve $C$, is the expectation 
value of the path ordered product of the gauge links along $C$. $W\left(R,T\right)$ denotes a rectangular Wilson loop with dimensions $R\times T$. There are two Feynman diagrams involving fermions, contributing to $W\left(R,T\right)$ at two loops, as shown in Fig. 1.

\begin{center}
\psfig{width=6truecm,file=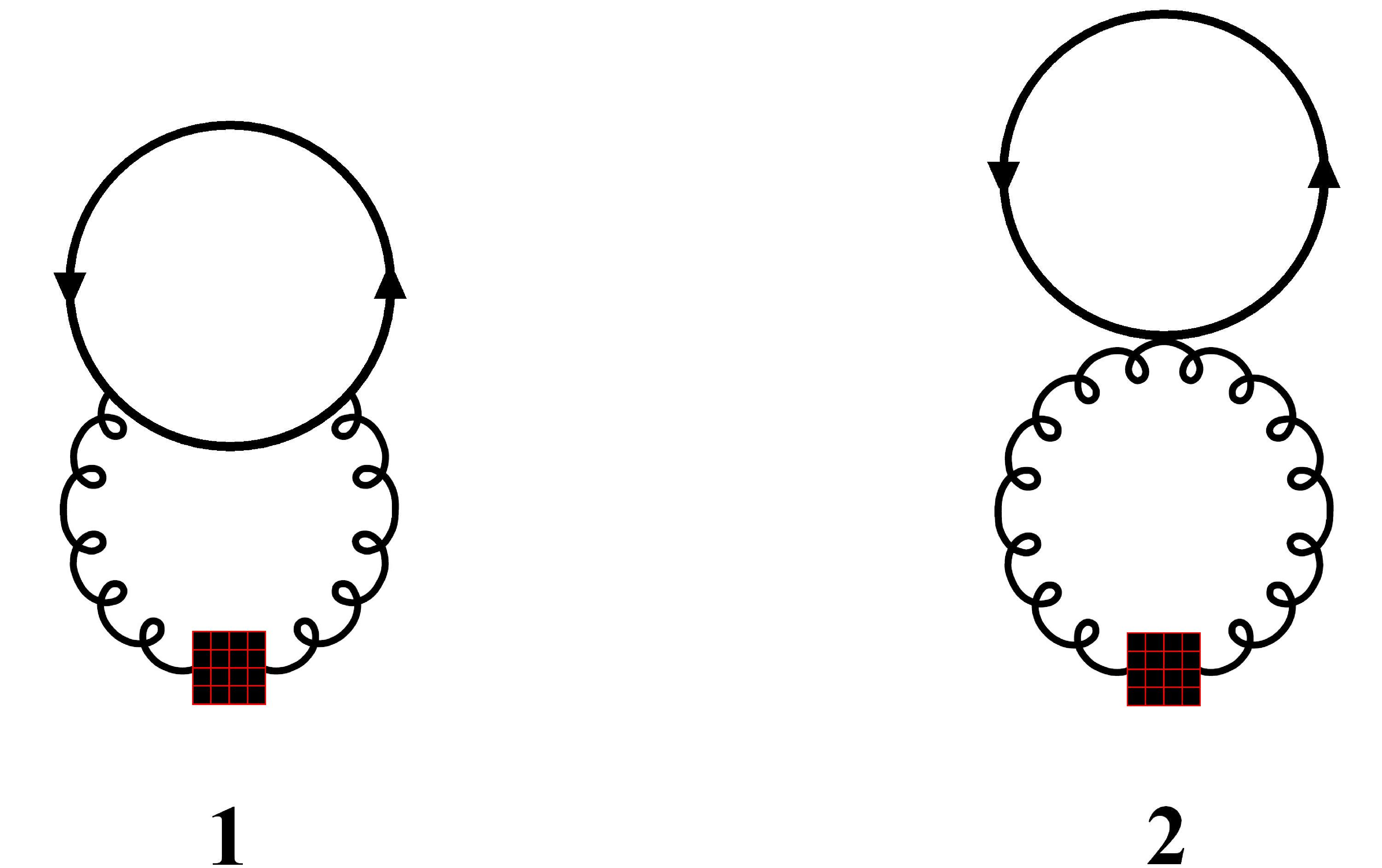}

\small
Fig.1: Two-loop fermionic diagrams contributing to $W\left(R,T\right)$.
\normalsize 
\end{center}

\noindent
The grid-like square stands for the 2-point vertex of $W\left(R,T\right)$, whose expression is:
\begin{eqnarray}
&W&\left(R,T\right)\rightarrow -{g^2\over24}\sum_{\mu, \nu}
\int {d^4kd^4k'\over\left(2 \pi\right)^4} A^a_{\mu}\left(k\right)A^b_{\nu}\left(k'\right)
\delta^{ab}\delta\left(k+k'\right) \Bigg[2\delta_{\mu,\nu}{\rm S} \left(k_{\mu}, R\right) \sum_{\rho}\sin^2\left(k_{\rho}aT/2\right)\nonumber \\
&+& 2\delta_{\mu,\nu} {\rm S} \left(k_{\mu}, T\right)\sum_{\rho}\sin^2\left(k_{\rho}aR/2\right) 
-4 {\rm S} \left(k_{\mu}, R\right) {\rm S} \left(k_{\nu}, T\right)\sin\left(k_{\mu}a/2\right)\sin\left(k_{\nu}a/2\right)\Bigg], 
\end{eqnarray}

\noindent
where: S$\left(k_{\mu}, R\right)\equiv{\displaystyle{\sin^2\left(Rk_{\mu}a/2\right)}/\displaystyle{\sin^2\left(k_{\mu}a/2\right)}}$.

The involved algebra of the lattice perturbation theory 
was carried out using our computer package in Mathematica. 
The value of each diagram is computed numerically for a 
sequence of finite lattice sizes. Their values have been summed, and then 
extrapolated to infinite lattice size.

\section{Calculation with Clover Fermions}

The fermionic part of the action contains an additional (clover) term, parameterized by a coefficient, $c_{SW}$, which is a free parameter in the present
work; $c_{SW}$ is normally tuned in a way as to minimize ${\cal O}(a)$ effects. The perturbative expansion of the Wilson loop is given 
by the expression:

\begin{equation}
W\left(C\right)\, /\, N=1-W_{LO}\ g^2-\left(W_{NLO}-{\left(N^2-1\right)\over N}N_f\Red{X}\right)g^4,
\label{Wc}
\end{equation}

\noindent
where: $W_{LO}$ and $W_{NLO}$ are the pure gauge contributions with the Wilson gauge action, which can be found in Ref.~\cite{SBPB,DRLS}, and:

\begin{equation}
\Red{X}=\Red{X_W}+\Red{X^a_{SW}}\Blue{c_{SW}}+\Red{X^b_{SW}}\Blue{c^2_{SW}}.
\label{Xi}
\end{equation}

We have computed the values of $\Red{X_W}$, $\Red{X^a_{SW}}$, and $\Red{X^b_{SW}}$. We compare our results with those of Ref.~\cite{DRLS} (only $X_{W}$)\footnote{In order to compare with Ref.~\cite{DRLS}, we deduce their values of $X_{W}$ from the data presented there and we estimate the errors stemming from that data.}, 
and Ref.~\cite{SBPB} (for $m=0$). Our results for $\Red{X_W}$, $\Red{X^a_{SW}}$, and $\Red{X^b_{SW}}$ as a 
function of mass for square loops, are shown in Figs. 2-4.

\begin{figure}[h]
\begin{minipage}{0.43\linewidth}
\begin{center}
\epsfig{file=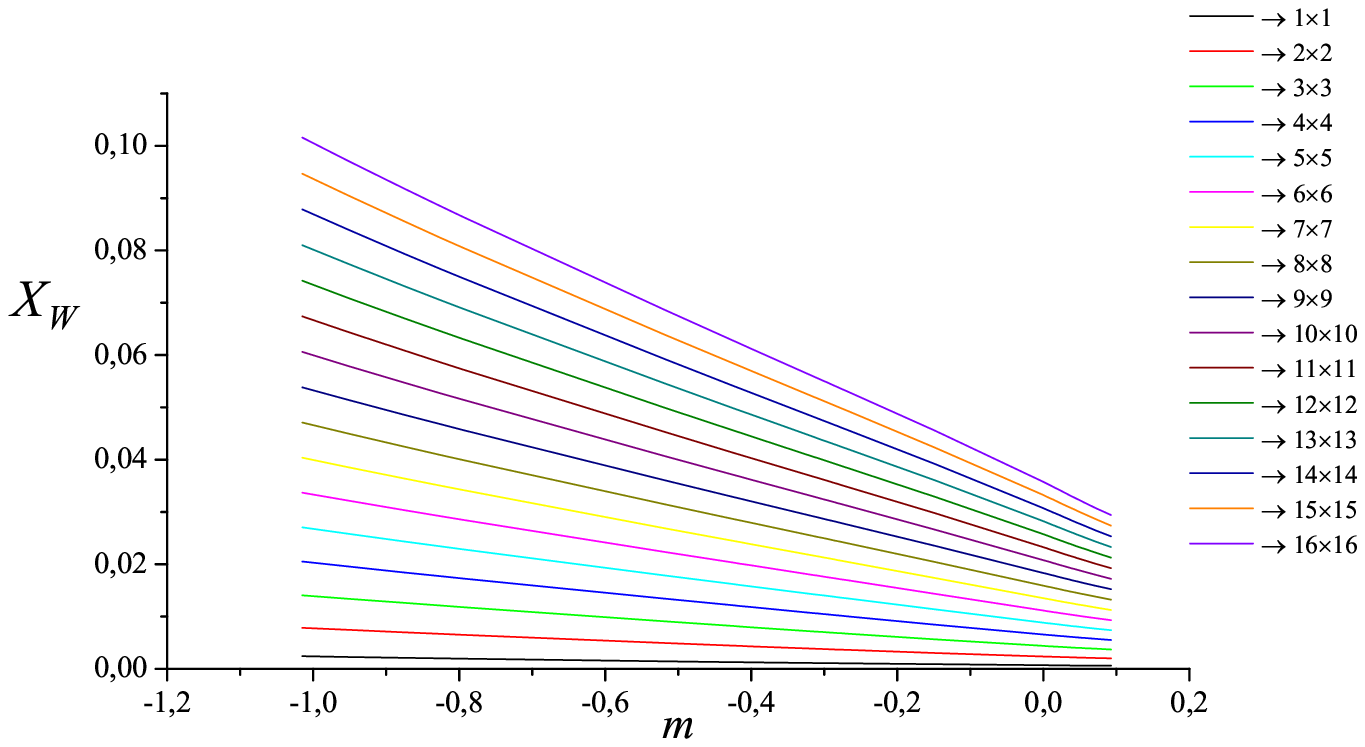, width=8truecm}

\small
Fig.2: $X_W$ for loops $L\times L$. Top line: $L=16$, bottom line: $L=1$.
\normalsize
\end{center}
\end{minipage}
\hfill
\begin{minipage}{0.43\linewidth}
\begin{center}
\epsfig{file=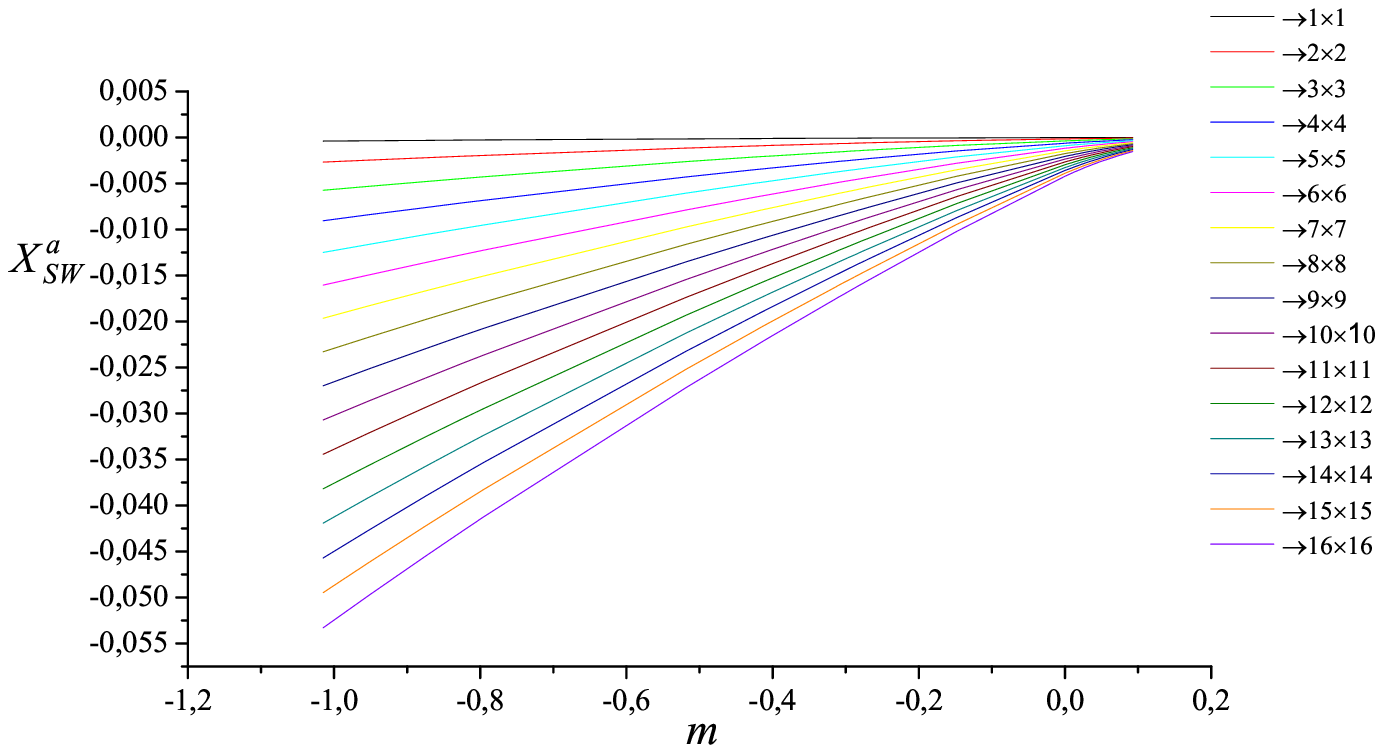, width=8truecm}

\small
Fig.3: $X^a_{SW}$ for loops $L\times L$. Top line: $L=1$, bottom line: $L=16$.
\normalsize
\end{center}
\end{minipage}
\end{figure}

\begin{figure}[h]
\begin{minipage}{0.43\linewidth}
\begin{center}
\epsfig{file=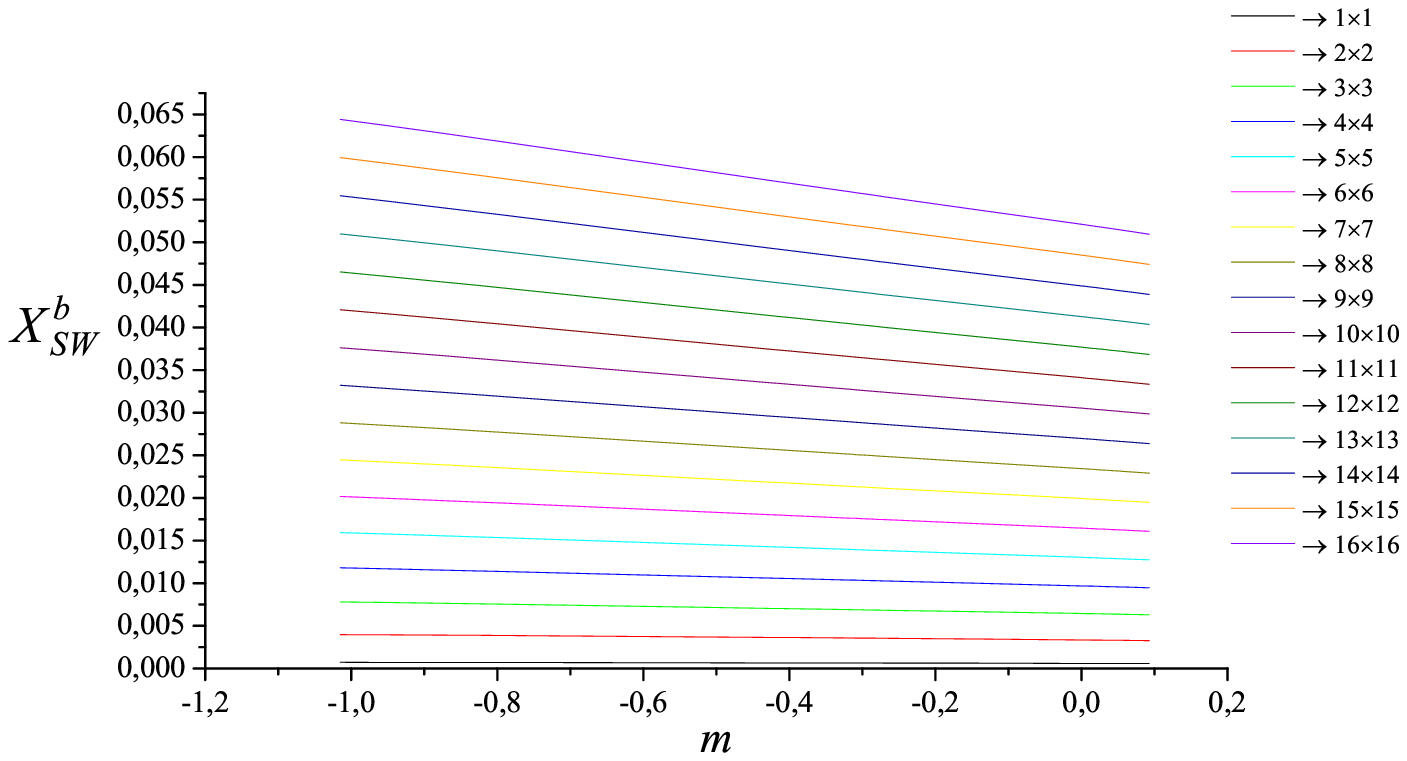, width=8truecm}

\small
Fig.4: $X^b_{SW}$ for loops $L\times L$. Top line: $L=16$, bottom line: $L=1$.
\normalsize
\end{center}
\end{minipage}
\hfill
\begin{minipage}{0.43\linewidth}
\begin{center}
\epsfig{file=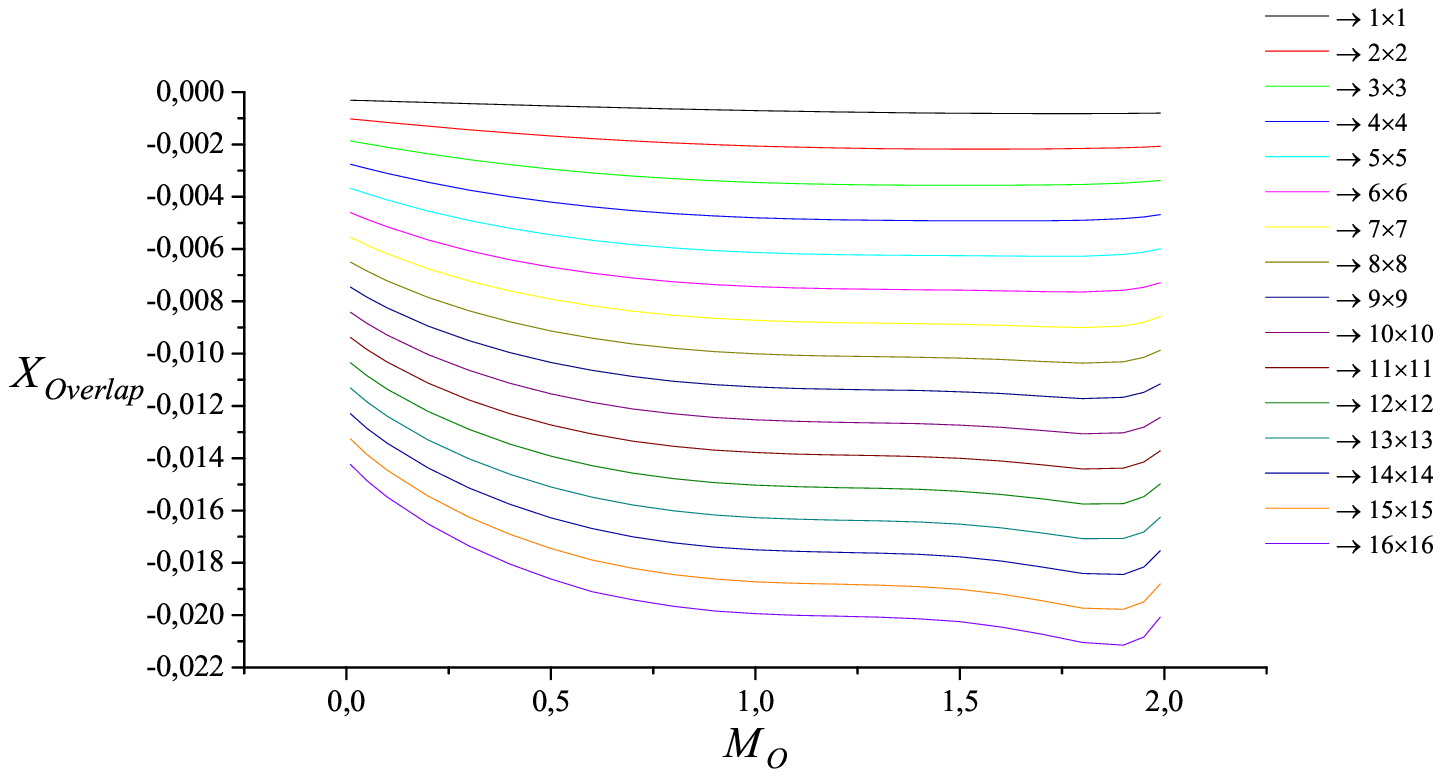, width=8truecm}

\small
Fig.5: $X_{Overlap}$ for loops $L\times L$. Top line: $L=1$, bottom line: $L=16$.
\normalsize
\end{center}
\end{minipage}
\end{figure}

\section{Calculation with Overlap Fermions}

The fermionic action now reads 

\begin{equation}
S_F =\sum_f \sum_{x,y} \bar{\psi}^f_x D_{\rm N}\left( x,y \right )\psi^f_y.
\label{latact}
\end{equation}

\noindent
with: $D_{\rm N} = M_O \left[1 + X (X^\dagger X)^{-1/2} \right]$,
and: $X = D_{\rm W} - M_O$; the index $f$ runs over $N_f$ flavours. 
Here, $D_{\rm W}$ is the massless
Wilson-Dirac operator with $r=1$, and $M_O$ is a free parameter whose
value must be in the range $0 <M_O < 2$, in order to guarantee the correct 
pole structure of $D_{\rm N}$.

Details on the propagator and vertices of $S_F$ can be found in our Ref.~\cite{AA} and references therein. The perturbative expansion of the Wilson loop is given by the expression:

\begin{equation}
W\left(C\right)\, /\, N=1-W_{LO}g^2-\left(W_{NLO}-{\left(N^2-1\right)\over N}N_f\Mulberry{X_{Overlap}}\right)g^4,
\label{Xoverlap}
\end{equation}
\noindent
where $\Mulberry{X_{Overlap}}$ are the values we compute. Fig. 5 shows $\Mulberry{X_{Overlap}}$ as a function of $M_O$ for square Wilson loops. Tables of our numerical values are presented in our Ref.~\cite{AAPH}.

\section{Calculation of the b-quark mass shift}

In perturbation theory, the expectation value of large Wilson loops decreases exponentially with the perimeter of the loops:

\begin{equation}
\left\langle W\left(R,T\right) \right\rangle \sim {\rm exp}\left(-2\delta m\left(R+T\right)\right).
\end{equation}

\noindent
Following Ref.~\cite{GMCTS}, the perturbative expansion 
for $\left\langle W\left(R,T\right) \right\rangle$ is:

\begin{equation}
\left\langle W\left(R,T\right) \right\rangle = 1 - g^2\, W_2\left(R,T\right)- g^4\, \Brown{W_4\left(R,T\right)}+ {\cal O}\left(g^6\right).
\end{equation}

\noindent
Using the expectation value of $W\left(R,T\right)$, we obtain the perturbative expansion for $\delta m$:

\begin{eqnarray}
\delta m = {1\over 2\left(R+T\right)}\Bigg[g^2\, W_2\left(R,T\right)
+ g^4\left(\Brown{W_4\left(R,T\right)}+{1\over2}W^2_2\left(R,T\right)\right)\Bigg]
\end{eqnarray}

\noindent
$W_2\left(R,T\right)$ involves only gluons and: $\Brown{W_4\left(R,T\right)}=\Brown{W^{g}_4\left(R,T\right)}+\Brown{W^{f}_4\left(R,T\right)}$, where $\Brown{W^{g}_4\left(R,T\right)}$ is the contribution in the pure gauge theory and $\Brown{W^{f}_4\left(R,T\right)}$ is the fermionic contribution.  

To evaluate the effect of fermions on the mass shift, we must examine their contribution in the limit as $R, T \rightarrow \infty$. To this end, we note that our expression assumes the generic form (modulo terms which will not contribute in this limit):
\begin{eqnarray}
\int{d^4p\over \left(2\pi\right)^4} \sin^2\left({p_{\nu}T/2}\right)\sin^2\left({p_{\mu}R/2}\right)\left({1\over \sin^2\left({p_{\nu}/2}\right)} +{1\over \sin^2\left({p_{\mu}/2}\right)}\right){1\over\left(\hat{p}^2\right)^2}G\left(p\right),
\label{Int}\end{eqnarray}

\noindent
where $\hat{p}^2=4\displaystyle{\sum_{\rho}} \sin^2\left(p_{\rho}/2\right)$ and $G\left(p\right)\sim p^2$. As $R, T \rightarrow \infty$, the above expression becomes:

\begin{eqnarray}
{1\over2}\left(R+T\right)\int {d^3\bar{p}\over \left(2\pi\right)^3}{1\over\left(\hat{\bar{p}}^2\right)^2}G\left(\bar{p}\right),
\end{eqnarray}

\noindent
where: $\bar{p}=\left(p_1,p_2,p_3,0\right)$ (for $\mu=4$ or $\nu=4$). 

The fermionic contribution takes the form:

\begin{eqnarray}
\Brown{W^{f}_4\left(R,T\right)} = \left(N^2-1\right)N_f\left(R+T\right)\Brown{V},
\end{eqnarray}

\noindent
where $\Brown{V}\equiv\Brown{V_W}+\Brown{V^a_{SW}}\Blue{c_{SW}}+\Brown{V^b_{SW}}\Blue{c^2_{SW}}$ for clover fermions, and $\Brown{V}\equiv\Brown{V_{Overlap}}$ for overlap fermions. The values of \Brown{$V_W$}, \Brown{$V^a_{SW}$}, \Brown{$V^b_{SW}$} and \Brown{$V_{Overlap}$} have been calculated in the present work. Figs. 6 and 7 show their values as a function of bare mass for clover fermions and of $M_O$ for overlap fermions:

\begin{figure} [h]
\begin{minipage}{0.43\linewidth}
\begin{center}
  \epsfig{file=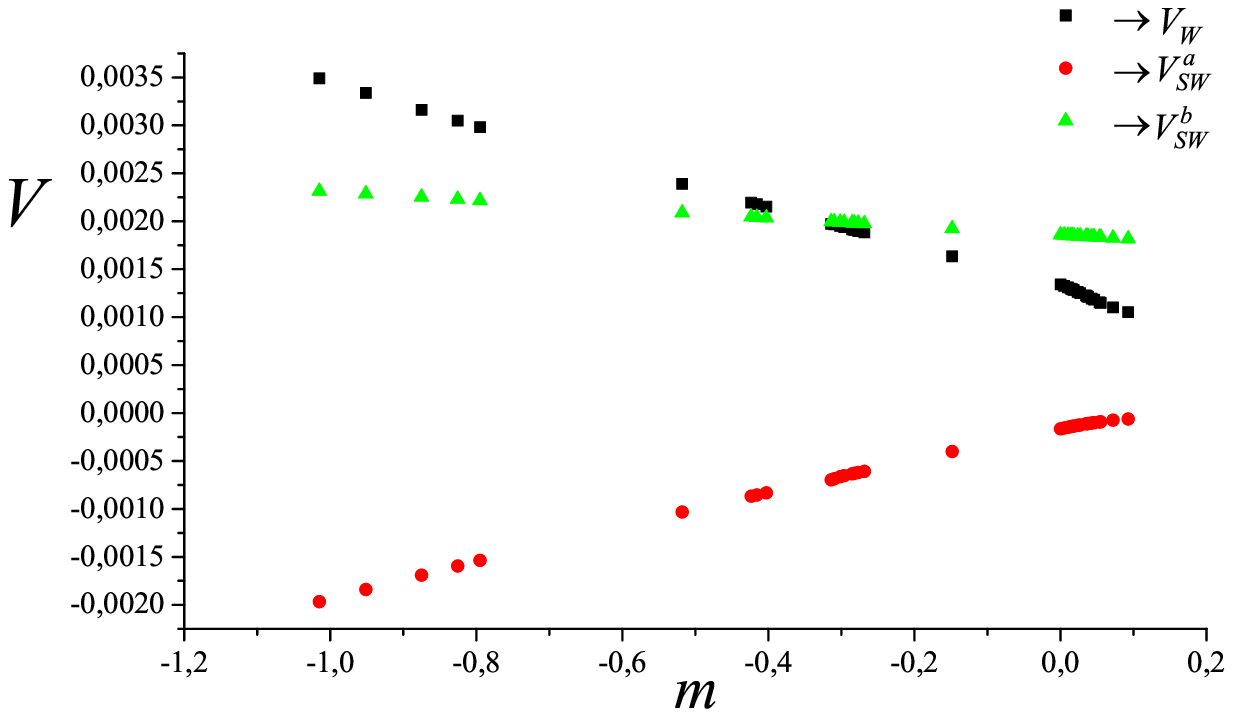, width=8truecm}

\small
Fig.6: $V_W$, $V^a_{SW}$ and $V^b_{SW}$ as a function of $m$. 
\normalsize
\end{center}
\end{minipage}
\hfill
\begin{minipage}{0.43\linewidth}
\begin{center}
  \epsfig{file=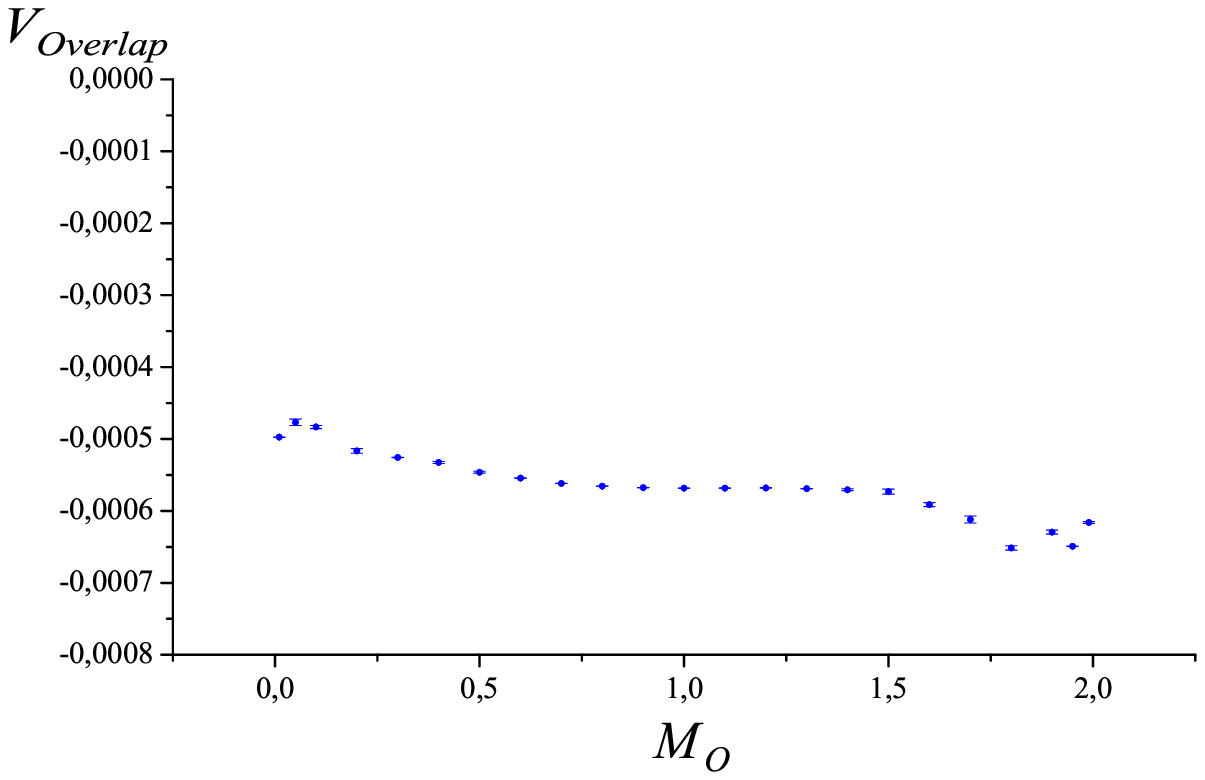, width=8truecm}

\small
Fig.7: $V_{Overlap}$ as a function of $M_O$.
\normalsize
\end{center}
\end{minipage}
\end{figure}

\noindent
At \MidnightBlue{one-loop} order the b-quark mass shift (for $N=3$) is given by:

\begin{equation}
a\delta m \simeq 0.16849 g^2 + {\cal{O}}\left( g^4\right),
\end{equation}

Using our results arrive at the \MidnightBlue{{\bf two-loop}} expression for $\delta m$. We list below some examples:

\noindent
The general form of $\delta m$ is ($\alpha_0=g^2/4\pi$):

\begin{eqnarray}
a \delta m \simeq 2.1173\alpha_0+\Bigg[11.152 - {\left(4\pi\right)^2\left(N^2-1\right)N_f \over 2 N}\Brown{V}\Bigg]\alpha_0^2+{\cal{O}}\left(\alpha_0^3\right).
\end{eqnarray}

For clover fermions, setting $m=0.0$, we find:

\begin{eqnarray}
\Brown{V}=0.00134096(5)- 0.0001641(1)\Blue{c_{SW}}+0.00185871(2)\Blue{c_{SW}^2}.
\end{eqnarray}

\noindent
These numbers agree with the result given in Ref.~\cite{GMCTS}, for the case $c_{SW}=0$, within the precision presented there. For \MidnightBlue{overlap} fermions, setting  $M_O=1.4$, our result is:

\begin{eqnarray}
\Brown{V_{Overlap}}=0.000571(1).
\end{eqnarray}

\section{Calculation of the Perturbative Static Potential}

The static potential is given by the expression:

\begin{eqnarray} 
aV\left(Ra\right) &=& -\lim_{T \rightarrow \infty} {d \ln W\left(R,T\right)\over dT} = V_1\left(R\right)g^2+V_2\left(R\right)g^4+\ldots,
\end{eqnarray}

\noindent
where: $V_1\left(R\right)$ is a pure gluonic contribution~\cite{SBPB}. 
$V_2\left(R\right)$ contains a gluonic part $V_g \left(R\right)$ which can be found in Ref.~\cite{SBPB} 
and a fermionic part \PineGreen{$F\left(R\right)$}:

\begin{eqnarray}
V_2\left(R\right) &=& V_g\left(R\right) -{\left(N^2-1\right)\over N}N_f\PineGreen{F\left(R\right)}.
\label{pot}
\end{eqnarray}

\noindent
We compute \PineGreen{$F\left(R\right)$} for clover and overlap fermions. For \MidnightBlue{clover}:

\begin{eqnarray}
\PineGreen{F\left(R\right)} = \PineGreen{F_{Clover}\left(R\right)}= F_W\left(R\right)+ F^a_{SW}\left(R\right)\Blue{c_{SW}}+ F^b_{SW}\left(R\right)\Blue{c^2_{SW}},
\end{eqnarray}

\noindent
and for \MidnightBlue{overlap}: \PineGreen{$F\left(R\right)} =\PineGreen{F_{Overlap}\left(R\right)$}. The values of $F_W\left(R\right)$, $F^a_{SW}\left(R\right)$, $ F^b_{SW}\left(R\right)$ and $F_{Overlap}\left(R\right)$ for specific mass values, can be found in the Tables of our Ref.~\cite{AAPH}. We also present them in Fig. 8 (for $c_{SW}=1.3$) and Fig. 9, as a function of $R$.

\begin{figure}[h]
\begin{minipage}{0.43\linewidth}
\begin{center}
  \epsfig{file=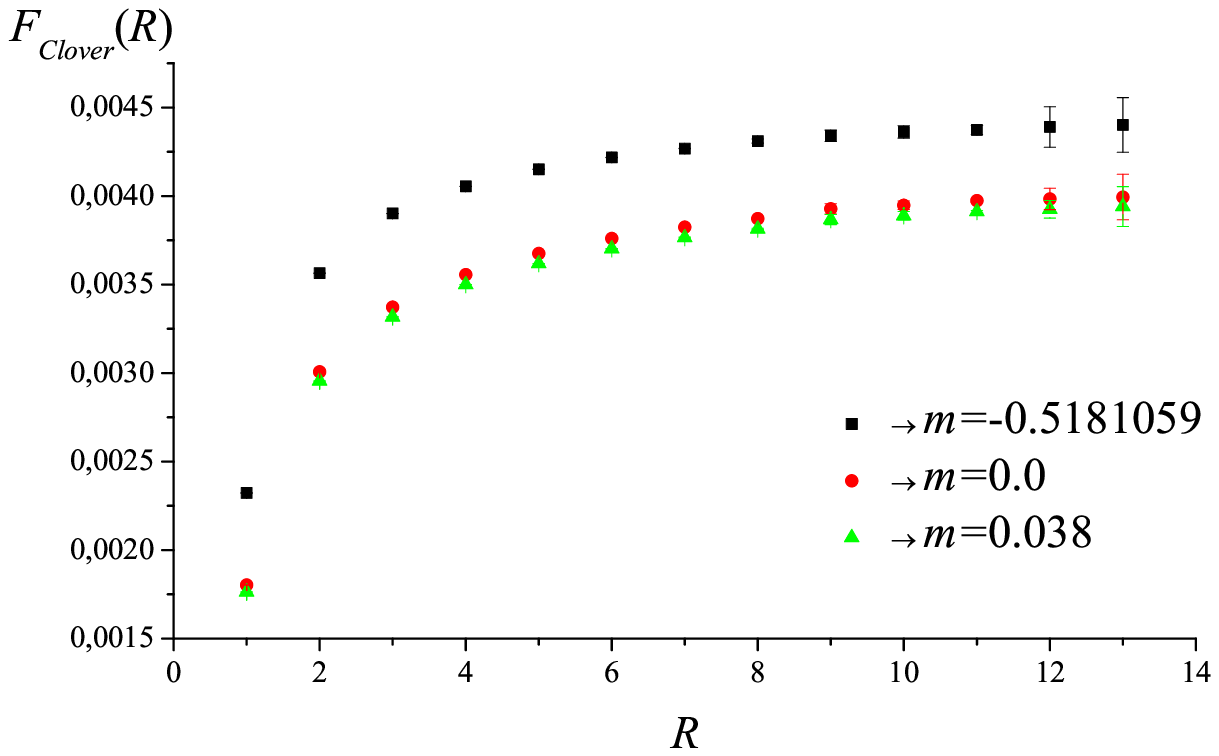,scale=0.5}
\small
Fig.8: $F_{Clover}$ as a function of $R$ ($c_{SW}=1.3$).
\normalsize  
\end{center}
\end{minipage} 
\hfill
\begin{minipage}{0.43\linewidth}
\begin{center}
  \epsfig{file=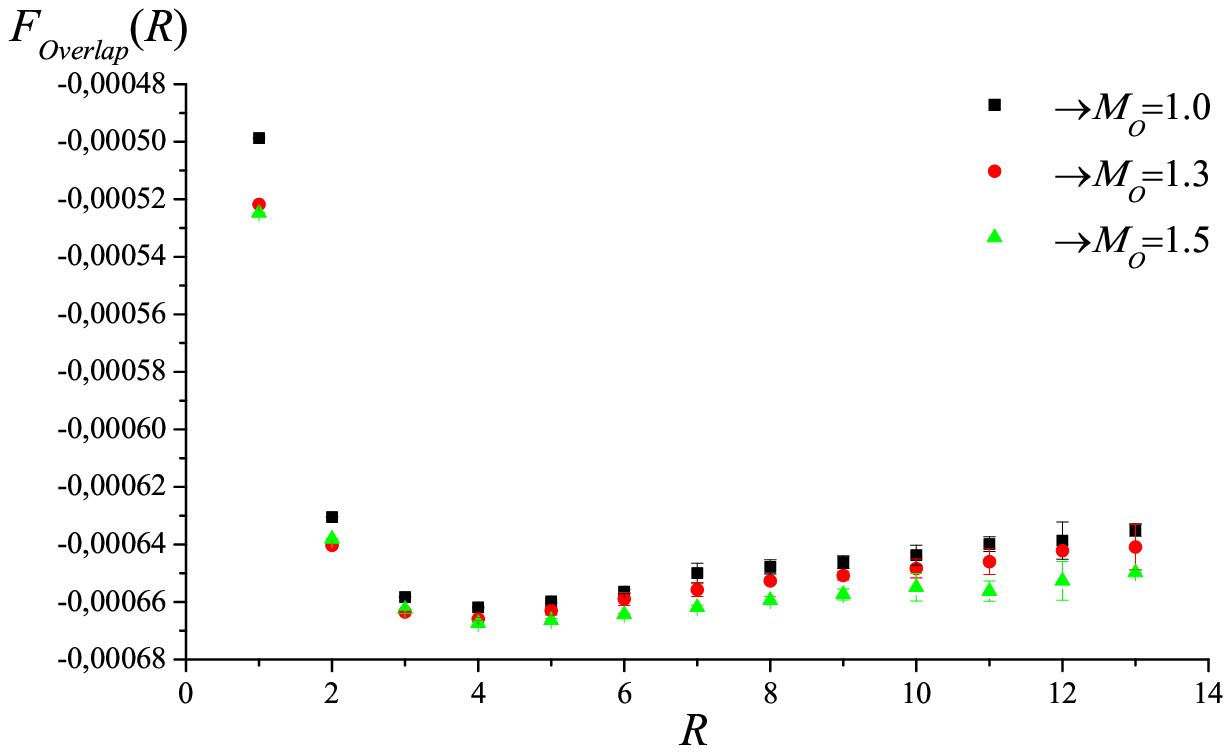,scale=0.5}
\small
Fig.9: $F_{Overlap}$ as a function of $R$.
\normalsize
\end{center}
\end{minipage}
\end{figure}

\vfill\eject

\begin{thebibliography}{99}

\bibitem{AAPH}A. Athenodorou and H. Panagopoulos, \emph{Large Wilson loops with overlap and clover fermions: Two-loop evaluation of the b-quark mass shift, and the quark-antiquark potential} [{\tt hep-lat/0509039}].

\bibitem{GMCTS} G. Martinelli and C. T. Sachrajda, \emph{Computation of the b-quark Mass with Perturbative Matching at the Next-to-Next-to-Leading Order}, \emph{Nucl. Phys.} {\bf B559} (1999) 429 [{\tt hep-lat/9812001}]. 

\bibitem{SBPB} G. Bali and P. Boyle, \emph{Perturbative Wilson loops with massive sea quarks on the lattice} [{\tt hep-lat/0210033}].

\bibitem{DRLS} F. Di Renzo and L. Scorzato, \emph{Numerical Stochastic Perturbation Theory for full QCD}, \emph{JHEP} {\bf 0410} (2004) 073 [{\tt hep-lat/0410010}].

\bibitem{AA} A. Athenodorou and H. Panagopoulos, \emph{The Lattice Free Energy with Overlap Fermions: A Two-Loop Result}, \emph{Phys. Rev.} {\bf D70} (2004) 077507 [{\tt hep-lat/0408020}].  

\end{thebibliography}
\end{document}